\documentclass[%
 reprint,
superscriptaddress,
 nofootinbib,
 aps,
 prd,
]{revtex4-2}

\usepackage{graphicx}
\usepackage{bm}
\usepackage[mathlines]{lineno}
\usepackage{amsmath,amssymb}
\usepackage{orcidlink}

\usepackage{tablefootnote,threeparttablex,booktabs}

\usepackage[colorinlistoftodos,prependcaption,textsize=tiny]{todonotes}

\usepackage{hyperref}
\usepackage[separate-uncertainty,retain-explicit-plus,per-mode=symbol,binary-units]{siunitx}
\usepackage[export]{adjustbox}
\usepackage{wrapfig}
\usepackage{ragged2e}
\usepackage{array,mathtools,dcolumn}
\usepackage{amsmath,amssymb}
\usepackage{wasysym}
\usepackage{stmaryrd}
\usepackage[below]{placeins}
\usepackage{soul}
\usepackage{tikz}
\usepackage{afterpage}
\usepackage{lineno}
\usepackage{listings}
\usepackage{array}
 \usepackage[normalem]{ulem}
 \useunder{\uline}{\ul}{}
\makeatletter
\@ifundefined{c@rownum}{%
  \let\c@rownum\rownum
}{}
\@ifundefined{therownum}{%
  \def\therownum{\@arabic\rownum}%
}{}
\makeatother
	\newcounter{rowno}
\usepackage{mhchem}
\usepackage{multirow}
\usepackage{eurosym}
\usepackage{pagecolor}
\usepackage{fancyhdr}
\usepackage{etoolbox}
\usepackage{calc}
\usepackage{dcolumn}
\usepackage{bm}
\usepackage{xargs}
\usepackage{xfrac}

\newcommandx{\unsure}[2][1=]{\todo[linecolor=red,backgroundcolor=red!25,bordercolor=red,#1]{#2}}
\newcommandx{\change}[2][1=]{\todo[linecolor=blue,backgroundcolor=blue!25,bordercolor=blue,#1]{#2}}
\newcommandx{\info}[2][1=]{\todo[linecolor=OliveGreen,backgroundcolor=OliveGreen!25,bordercolor=OliveGreen,#1]{#2}}
\newcommandx{\improvement}[2][1=]{\todo[linecolor=Plum,backgroundcolor=Plum!25,bordercolor=Plum,#1]{#2}}
\newcommandx{\thiswillnotshow}[2][1=]{\todo[disable,#1]{#2}}

\newcommand{\refcite}[1]{Ref.~\cite{#1}}
\newcommand{\refscite}[1]{Refs.~\cite{#1}}
\newcommand{\reffig}[1]{Fig.~\ref{#1}}

\newcommand{\reffiginit}[1]{Figure~\ref{#1}}
\newcommand{\reftab}[1]{Table~\ref{#1}}
\newcommand{\refeqn}[1]{Eq.~(\ref{#1})}

\usepackage{paralist}
\usepackage{tcolorbox}
\usepackage{pgfornament}
\usepackage{array}
\usepackage[nodayofweek]{datetime}
\usepackage{import}
\usepackage{appendix}
\usepackage{cleveref}
\usepackage{enumitem}
\setdescription{itemsep=0pt,parsep=0pt,labelsep=.2cm,leftmargin=2.2cm,labelwidth=2cm,listparindent=.7cm}
\setlist{nolistsep,leftmargin=1cm}
\newlist{enumcompactitem}{itemize}{3}
\setlist[enumcompactitem]{topsep=0pt,partopsep=0pt,itemsep=0pt,parsep=0pt}
\setlist[enumcompactitem,1]{label=\textbullet}
\setlist[enumcompactitem,2]{label=---}
\setlist[enumcompactitem,3]{label=*}
\newlist{enumcompactdesc}{description}{3}
\setlist[enumcompactdesc]{topsep=0pt,partopsep=0pt,itemsep=0pt,parsep=0pt}
\newlist{enumcompactenum}{enumerate}{3}
\setlist[enumcompactenum]{topsep=0pt,partopsep=0pt,itemsep=0pt,parsep=0pt}
\setlist[enumcompactenum,1]{label=\arabic*}
\setlist[enumcompactenum,2]{label=\alph*}
\setlist[enumcompactenum,3]{label=\roman*}
\newcommand{\isotope}[2]{\mbox{$^{#1}$#2}}

\newcommand{\insitu}{\emph{in situ}}

\DeclareSIUnit\c{\mbox{$c$}}
\DeclareSIUnit\magn{\mbox{$\times$}}
\DeclareSIUnit\min{min}
\DeclareSIUnit\hr{hr}
\DeclareSIUnit\hrs{hrs}
\DeclareSIUnit\week{week}
\DeclareSIUnit\month{mo}
\DeclareSIUnit\months{mos}
\DeclareSIUnit\year{yr}
\DeclareSIUnit\years{years}
\DeclareSIUnit\yr{yr}
\DeclareSIUnit\standard{std}
\DeclareSIUnit\str{sr}
\DeclareSIUnit\ppm{ppm}
\DeclareSIUnit\ppb{ppb}
\DeclareSIUnit\ppt{ppt}
\DeclareSIUnit\pe{PE}
\DeclareSIUnit\spe{SPE}
\DeclareSIUnit\pdm{PDM}
\DeclareSIUnit\ev{events}
\DeclareSIUnit\ct{counts}
\DeclareSIUnit\neutron{\mbox{$n$}}
\DeclareSIUnit\smp{samples}
\DeclareSIUnit\Sample{S}
\DeclareSIUnit\ch{ch}
\DeclareSIUnit\hit{hit}
\DeclareSIUnit\hits{hits}
\DeclareSIUnit\bin{(\mbox{5-PE}~bin)}
\DeclareSIUnit\sgm{\mbox{$\sigma$}}
\DeclareSIUnit\rms{RMS}
\DeclareSIUnit\keVee{\mbox{keV$_{{\rm ee}}$}}
\DeclareSIUnit\keVr{\mbox{keV$_{\rm nr}$}}
\DeclareSIUnit\eVee{\mbox{eV$_{\rm ee}$}}
\DeclareSIUnit\eVr{\mbox{eV$_{\rm nr}$}}
\DeclareSIUnit\ph{photon}
\DeclareSIUnit\el{\mbox{$e^-$}}
\DeclareSIUnit\pm{\mbox{PMT}}
\DeclareSIUnit\pixel{\mbox{pixel}}
\DeclareSIUnit\inch{''}
\DeclareSIUnit\foot{'}
\DeclareSIUnit\bit{bit}
\DeclareSIUnit\sample{samples}
\DeclareSIUnit\barn{barn}
\DeclareSIUnit\bara{bar}
\DeclareSIUnit\bar{bar}
\DeclareSIUnit\barg{barg}
\DeclareSIUnit\mlardepth{\mbox(meter~of~\LAr~depth)}
\DeclareSIUnit\Curie{Ci}
\DeclareSIUnit\psia{psia}
\DeclareSIUnit\psf{psf}
\DeclareSIUnit\pcf{pcf}
\DeclareSIUnit\parsec{pc}
\DeclareSIUnit\cps{cps}
\DeclareSIUnit\mwe{\mbox{m.w.e.}}
\DeclareSIUnit\liveday{\mbox{live-days}}
\DeclareSIUnit\days{\mbox{days}}
\DeclareSIUnit\miles{\mbox{miles}}
\DeclareSIUnit\lumens{\mbox{lm}}
\DeclareSIUnit\degreeC{\mbox{$^{\circ}$C}}
\DeclareSIUnit\degreeF{\mbox{$^{\circ}$F}}
\DeclareSIUnit\electron{\mbox{$e^-$}}
\DeclareSIUnit\Euro{\mbox{\euro}}
\DeclareSIUnit\cph{cph}
\DeclareSIUnit\neq{neq}
\DeclareSIUnit\normal{\mbox{N}}
\DeclareSIUnit\USD{\mbox{\$}}

\newcommand{\MnFiveFourHalfLife}{\SI{312.1}{\day}}

\newcommand{\CoSixZeroHalfLife}{\SI{5.27}{\year}}

\newcommand{\KrEightFiveHalfLife}{\SI{10.8}{\year}}

\newcommand{\DSf}{\mbox{DarkSide-50}}

\newcommand{\LAr}{\ce{LAr}}

\newcommand{\pe}{\mbox{PE}}

\def\apjs{ApJS}%

\begin{document}

\preprint{APS/123-QED}

\title{Search for dark matter annual modulation with  DarkSide-50}

\author{P.~Agnes}\affiliation{Department of Physics, Royal Holloway University of London, Egham TW20 0EX, UK}
\author{I.F.M.~Albuquerque}\affiliation{Instituto de F\'isica, Universidade de S\~ao Paulo, S\~ao Paulo 05508-090, Brazil}
\author{T.~Alexander}\affiliation{Pacific Northwest National Laboratory, Richland, WA 99352, USA}
\author{A.K.~Alton}\affiliation{Physics Department, Augustana University, Sioux Falls, SD 57197, USA}
\author{M.~Ave}\affiliation{Instituto de F\'isica, Universidade de S\~ao Paulo, S\~ao Paulo 05508-090, Brazil}
\author{H.O.~Back}\affiliation{Pacific Northwest National Laboratory, Richland, WA 99352, USA}
\author{G.~Batignani}\affiliation{INFN Pisa, Pisa 56127, Italy}\affiliation{Physics Department, Universit\`a degli Studi di Pisa, Pisa 56127, Italy}
\author{K.~Biery}\affiliation{Fermi National Accelerator Laboratory, Batavia, IL 60510, USA}
\author{V.~Bocci}\affiliation{INFN Sezione di Roma, Roma 00185, Italy}
\author{W.M.~Bonivento}\affiliation{INFN Cagliari, Cagliari 09042, Italy}
% ...
\collaboration{The DarkSide-50 Collaboration}\noaffiliation

\date{\today}

\begin{abstract}
Dark matter may induce an event in an Earth-based detector, and its event rate is predicted to show an annual modulation as a result of the Earth's orbital motion around the Sun.
We searched for this modulation signature using the ionization signal of the \DSf\ liquid argon time projection chamber.
No significant signature compatible with dark matter is observed in the electron recoil equivalent energy range above \SI{40}{\eVee}, the lowest threshold ever achieved in such a search. 
\end{abstract}

\maketitle


\section{Introduction}
The combined effect of Earth's rotations around the Sun and the Galactic Center is expected to produce an annual modulation of the dark matter particle interaction rate in terrestrial detectors~\cite{Drukier:1986tm}, thereby offering a unique signature for directly probing dark matter particles and unveiling their true nature.
The DAMA/LIBRA experiment claimed the detection of such a signature in their NaI detectors in the \si{\keV} range~\cite{Bernabei:2013xsa,Bernabei:2021kdo}.
The interpretation of this claim with the weakly interacted massive particle hypothesis (WIMP) is currently facing challenges due to the null detection of WIMP-induced nuclear-recoil signals in other experiments~\cite{EDELWEISS:2016nzl,LUX:2016ggv,SuperCDMS:2017mbc,XENON:2018voc,DarkSide:2018kuk,DEAP:2019yzn,PICO:2019vsc,CRESST:2019jnq,COSINE-100:2021xqn,XMASS:2022tkr,LZ:2022lsv,XENON:2023cxc}.
Several experiments, such as ANAIS-112~\cite{Amare:2021yyu} and COSINE-100~\cite{COSINE-100:2021zqh}, have been making progress toward a model-independent test of the DAMA/LIBRA's claim adopting NaI detectors.
Another approach to test this claim and possibly to reveal WIMP properties can be offered by searching for the modulation with other detectors which have different target materials, background sources, energy resolution, and experimental sites.
Such results from xenon-based dark matter experiments are reported by XENON-100~\cite{XENON:2017nik}, LUX~\cite{LUX:2018xvj}, and XMASS~\cite{XMASS:2022tkr} collaborations, though none of them have confirmed the positive claim above \SI{1}{\keV} electron recoil equivalent~(\SI{}{\keVee}).

Dual-phase noble-liquid time projection chambers (TPCs) measure the scintillation and ionization signals from a particle interacting in the liquid.
Such detectors were originally designed to discover and have led the search for WIMPs with masses above \SI{10}{\giga\eV\per\square\c}.
Moreover, in the last decade, they have also exhibited world-class sensitivity to light dark matter candidates exploiting only the ionization signal spectrum above a few detected ionization electrons (\(N_e\))~\cite{XENON10:2011prx,Essig:2012yx,Essig:2017kqs,XENON:2016jmt,DarkSide:2018bpj,DarkSide:2018ppu,XENON:2019gfn,XENON:2019zpr,PandaX-II:2021nsg}.
Among them, the \DSf\ detector, a liquid argon (LAr) TPC located underground at the Laboratori Nazionali del Gran Sasso (LNGS)~\cite{DarkSide:2014llq,DarkSide:2015cqb,DarkSide:2018kuk}, recently demonstrated an unprecedented sensitivity in this energy region~\cite{DarkSide-50:2022qzh,DarkSide:2022dhx,DarkSide:2022knj,DarkSide-50:2023fcw}.
This achievement was accomplished by looking for an event excess in the energy spectrum with respect to the background model above \SI{0.06}{\keVee}.
In this work, we report for the first time on the search for the annual rate modulation of events down to \SI{0.04}{\keVee}, the lowest threshold ever achieved in a dark matter modulation search.
The analysis relies on two approaches: the maximum likelihood fit and the Lomb-Scargle periodogram~\cite{2018ApJS..236...16V}.
The results are also compared to the claim by the DAMA/LIBRA experiment assuming that the dark matter produces signals of the same electron-recoil-equivalent-energy in both NaI and LAr detectors.

\section{Detector}
The DarkSide-50 detector and associated apparatus are described in detail in \refscite{DarkSide:2014llq,DarkSide:2017odo,DarkSide-50:2023nes}.
Here we give a brief overview of the experimental apparatus.

DarkSide-50 consists of three nested detector systems, the LAr TPC, the neutron veto, and the cosmic muon veto.
The TPC contains an active liquid target of \SI{46.4\pm0.7}{\kilo\gram}.
It is housed in a stainless steel double-walled, vacuum-insulated cryostat, shielded by a \SI{30}{\tonne} boron-loaded liquid scintillator veto instrumented with 110 8-inch PMTs. 
The purpose of this is to actively tag neutrons \insitu. 
A \SI{1}{\kilo\tonne} water \v Cerenkov veto, equipped with 80 PMTs, surrounds the neutron veto to actively tag cosmic muons and to passively shield the TPC against external backgrounds~\cite{DarkSide:2015fzz}.

Two arrays of 19 3-inch photomultiplier tubes (PMTs), located at the top and the bottom of the TPC, detect light pulses from scintillation (S1) induced by particle interactions in the liquid bulk. 
The same interactions generate ionization electrons, which are drifted through the LAr volume by a \SI{200}{\volt\per\centi\meter} electric field up to the top of the TPC. 
Then, they are extracted into the gas phase by a \SI{2.8}{\kilo\volt\per\centi\meter} field and induce delayed photon pulses (S2) by electroluminescence under a \SI{4.2}{\kilo\volt\per\centi\meter} field, as characterized in \refcite{DarkSide:2018stg}.

\DSf\ started taking data in April 2015 with a low-radioactivity LAr target, extracted from a deep underground source (UAr)~\cite{DarkSide:2015cqb}, and concluded the operations in February 2018.
We do not use a short period of time in July 2015 in which the inline argon purification getter was bypassed and an enhanced event rate was observed near the analysis threshold~\cite{DarkSide:2018bpj}.
In addition, the first four months of data were contaminated by the cosmogenic \isotope{37}{Ar} isotope, with a half-life of \SI{35.0}{\day}~\cite{be:cea-02476417}, and were only used to calibrate the ionization response~\cite{DarkSide:2021bnz}. 
About 25\% of the rest of the data taking was devoted to calibration campaigns with dissolved and external radioactive sources. 
The livetime used in this paper corresponds to \SI{693.3}{\day}.

\section{Analysis}
\subsection{Dataset}
The data used in this analysis is acquired upon a hardware event trigger requiring a coincidence of two or more PMT signals above \SI{0.6}{photoelectron} within \SI{100}{\nano\second}~\cite{DarkSide:2017odo}.
Selected events for further analysis in this dataset are required to be single-scatter, i.e., with a single S2 pulse.
These events must also be isolated in time from the preceding events, following a veto of \SI{20}{\milli\second} after any event triggering the data acquisition system.
Additional cuts are used to remove pile-up pulses, which are too close in time such that the pulse finder algorithm is unable to separate the clusters, and surface $\alpha$ events, characterized by a large S1 plus an anomalously low S2 because of absorption of the ionization electrons into the detector wall.
Finally, we remove events reconstructed in the outer $\sim$\SI{7}{\centi\meter} thick cylindrical shell of the TPC, resulting in the \SI{19.4}{\kilo\gram} fiducial volume in the center.
The low energy threshold for this analysis is defined in order to reject spurious electrons~(SEs)~\cite{DarkSide:2018bpj,DarkSide-50:2022qzh}.
These are considered to originate from ionization electrons trapped on impurities along the drift in LAr, and released with a certain delay, as will be the object of a paper in preparation.
A full description of the selection criteria can be found in \refcite{DarkSide-50:2022qzh}.

\subsection{Background model}
The time evolution of background events can be described by the combination of a set of decaying exponentials and a constant term.
The latter component includes the radioactive backgrounds whose lifetime is much longer than the data-taking period of about three years and is dominated by the $\beta$-decay of \isotope{39}{Ar} (\SI{268}{\year}~\cite{Chen:2018trb}).
The exponential components arise from the decays of \isotope{37}{Ar} (\SI{35.0}{\day}~\cite{be:cea-02476417}), \isotope{85}{Kr} (\KrEightFiveHalfLife~\cite{be:cea-02476107}), \isotope{54}{Mn} (\MnFiveFourHalfLife~\cite{be:cea-02476107}), and \isotope{60}{Co} (\CoSixZeroHalfLife~\cite{be:cea-02476243}). 
The first two isotopes are intrinsically present in LAr, while the latter two are contaminants of the PMTs, and \isotope{60}{Co} is also present in the cryostat stainless steel. 
The latter two emit \(\gamma\)- and x-rays, which deposit energy in the LAr target. 
The background model is generated with the DarkSide-50 \texttt{Geant4}-based Monte Carlo~\cite{DarkSide:2017wdu} code.
The model is built on data from an extensive material screening campaign to characterize the trace radioactivity content of every detector component.
It also uses \insitu\ measurements with \DSf~\cite{DarkSide-50:2022qzh} and incorporates the detector response model~\cite{DarkSide:2021bnz}.

\subsection{Detector stability}
A crucial aspect for this analysis is the long-term stability of the detector performance, monitored by various sensors incorporated inside the cryogenic system, as well as by the recorded events from the TPC itself. 
The two parameters whose fluctuations may potentially have a high impact on the modulation search are the electric drift field, \(F\), and the average number of detected S2 photons per ionization electron extracted in the gas phase, \(g_2\).
The stability of \(F\) is monitored \insitu\ via the stability of the edge of the drift-time distribution that corresponds to the very bottom of the TPC~\cite{DarkSide:2018stg}. This is allowed by the fact that a large part of the events in \DSf\ come from the diffused isotopes of \isotope{39}{Ar} and \isotope{85}{Kr}.
The maximum fluctuation of \(F\) was estimated to be less than 0.01\%, too small to affect the ionization response.
Based on the S2/S1 ratio for electronic recoil events above the region of interest (RoI) ([0.04, 20.0]~\si{\keVee}), \(g_2\) varies at most by 0.5\% over the whole data-taking period.
The impact on the modulation signal searches described later from the measured instability is evaluated by pseudoexperiments.
It is found that any possible bias on the result is smaller than the size of statistical fluctuations.

We also check the temporal evolution of other detector parameters, such as the liquid argon purity, pressure and temperature of the gaseous argon, PMT response to single photoelectron, and the condition of the inline filters to maintain pure argon.
A systematic study on the stability of such parameters can be found in~\refcite{DarkSide-50:2023nes}.
Throughout the work we find that the stabilities of most parameters are typically $\mathcal{O}(0.1\%)$ or less such that they do not affect the observed event rate.
An exception is the liquid argon purity which continuously increases from \ce{O2} equivalent contamination of \SI{60}{\ppt} (corresponding to the drift electron lifetime of \SI{5}{\milli\second}) to $<$\SI{15}{\ppt}~($>$\SI{20}{\milli\second}).
A toy Monte Carlo study shows that such an increase cannot make any fake modulation signal, as the maximum electron drift time (\SI{376}{\micro\second}) is much shorter than that level.
Another exception is the temperature of a charcoal trap for radon removal which is put inside the gas circulation line to maintain purified liquid argon.
The instability is observed at the level of 1\%.
Since we do not see any correlation between the temperature and the observed event rate in various energy ranges, and we do not find any way for the instability to affect the TPC observed event rate, we affirm that it does not influence the following analysis.


\section{Result\label{sec:result}}
\subsection{Phase-free likelihood fit}
We first perform a likelihood fit to search for annual modulation signal without constraining its phase.
Since the observed events below \SI{4}{\el} are contaminated by the SE background~\cite{DarkSide-50:2022qzh}, of which \textit{a priori} expectation is still missing, we define two ranges to be analyzed as [4, 41]~\si{\el} and [41,68]~\si{\el} ranges, corresponding to [0.06,2.0]~\si{\keVee} and [2.0,6.0]~\si{\keVee}, respectively.

\begin{figure}[t]
    \centering
    \includegraphics[width=\linewidth]{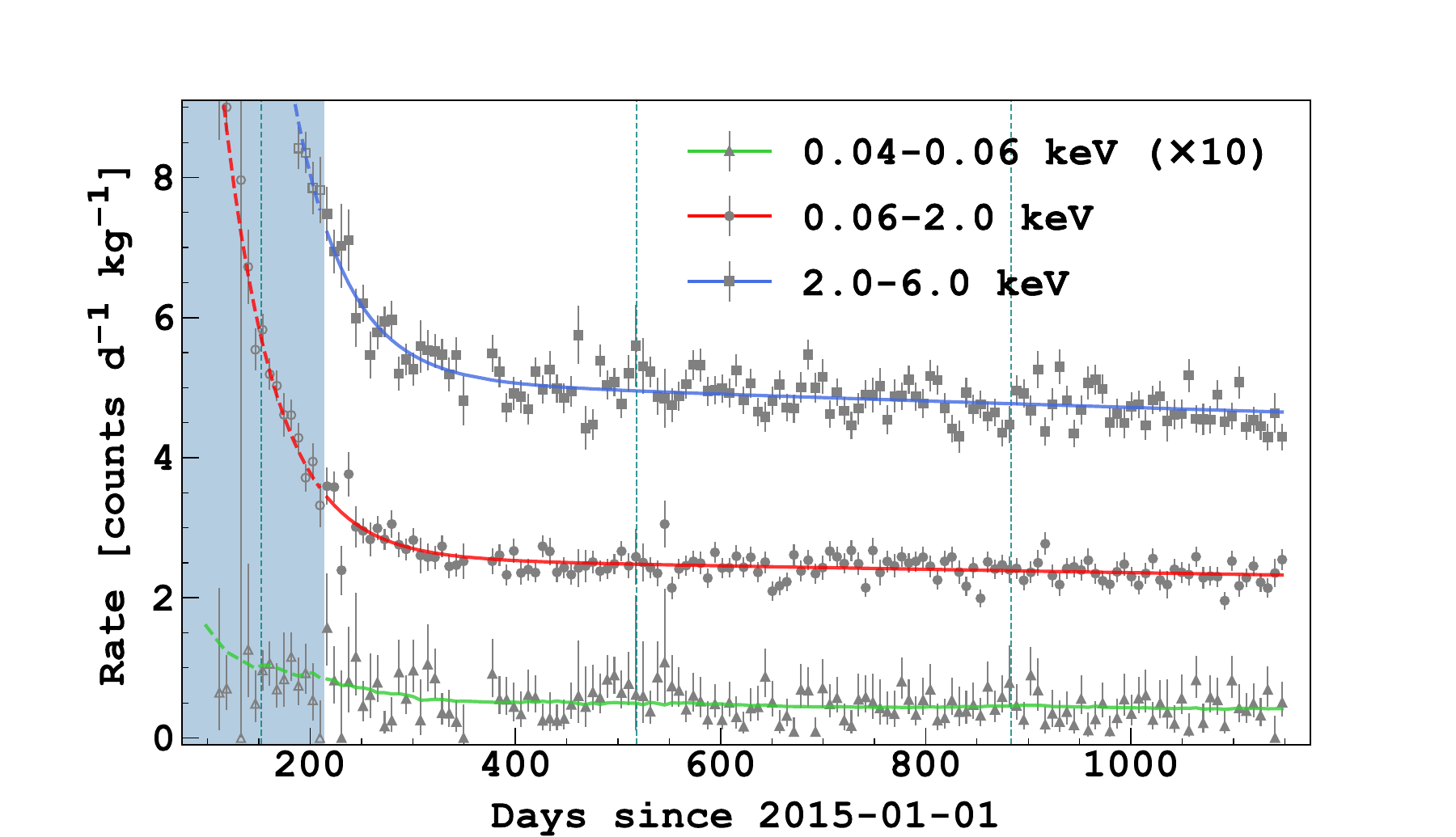}
    \caption{Temporal evolution of the observed event rates for [3, 4]~\si{\el} (corresponding to [0.04, 0.06]~\si{\keVee}), [4, 41]~\si{\el} ([0.06, 2.0]~\si{\keVee}), and [41, 68]~\si{\el} ([2.0, 6.0]~\si{\keVee}) ranges.
    The bin width is \SI{7}{\day}.
    The colored solid lines represent the background-only fit.
    The vertical dotted lines correspond to June 2nd, which is when the dark matter induced event rate has its maximum.
    The blue-shaded region corresponds to the first four months devoted to the detector calibration and is thus excluded from this analysis.
    }
    \label{fig:rate_bestfit}
\end{figure}

\reffiginit{fig:rate_bestfit} shows the measured time-dependent event rates for events with \(N_e\) in the [4, 41]~\si{\el} and [41,68]~\SI{}{\el} ranges.
The signal and backgrounds are modeled with 
\begin{equation}
    \label{eq:pdf}
    f(t) = A_\chi \cos\Bigl(\frac{t-\phi}{T/2\pi}\Bigr) +
    \sum_{l} \frac{A_{l}}{\tau_{l}} e^{-t/\tau_{l}} + C,     
\end{equation}
where $l = (^{37}\mathrm{Ar}, ^{85}\mathrm{Kr}, ^{54}\mathrm{Mn}, ^{60}\mathrm{Co})$, $A_\chi$ is the amplitude of the modulated term of the signal, $\phi$ the phase, and $T$ the period fixed to \SI{1}{\year}. 
The constant term $C$ is the sum of the time-averaged signal component and long-lived backgrounds. 
The parameters $\tau_{l}$ and $A_{l}$ correspond to the decay times and amplitudes, respectively, of the short-lived isotopes \(l\). 
The background-only fits to data, by fixing $A_\chi=0$, are shown in \reffig{fig:rate_bestfit} for the two ranges. 

The statistical significance of a possible modulated signal is assessed using the following binned likelihood with the bin width of \SI{7}{\day},
\begin{multline}
  \mathcal{L}  =   \prod_{i\,\in\,t_\textrm{bins}} \mathcal{P} \left( n_i| m_i(A_\chi, \phi,C,\Theta)\right) \\
  \times  \prod_{\theta_k\,\in\, \Theta}\mathcal{G}(\theta_k|\theta^0_k\textrm{,}\,\Delta \theta_k).
\label{eq:likelihood}
\end{multline}
The first term represents the Poisson probability of observing $n_i$ events in the $i^{th}$ time bin with respect to the expected ones, $m_i(A_\chi, \phi, C, \Theta)$, evaluated with \refeqn{eq:pdf}. 
In the fit, \(A_\chi\), \(\phi\) and \(C\) are left free to vary, while the other parameters are contained inside \(\Theta\), which represents the set of remaining nuisance parameters constrained by the Gaussian penalty terms in the last factor of \refeqn{eq:likelihood}. 
In the latter, \(\theta^0_k\) and \(\Delta \theta_k\) represent the nominal central values and uncertainties, respectively, of the nuisance parameters and are listed in \reftab{tab:nuis_par}.
The nuisance parameters account for uncertainties on the fiducial volume of the TPC (which induces a 1.1\% uncertainty on the event rate from \isotope{54}{Mn} and \isotope{60}{Co} in the PMTs and cryostat; and a 1.5\% uncertainty on the other event rates, acting in a correlated way~\cite{DarkSide-50:2022qzh}) and on the activities of short-lived decays in the energy range of interest. 
These are obtained from the combination of the uncertainty on the measured rate (14\%, 4.7\%, 40\%, 12\% for $^{37}$Ar, $^{85}$Kr, $^{54}$Mn, $^{60}$Co, respectively~\cite{DarkSide-50:2022qzh}),  with the uncertainty arising from the definition of the energy range due to the ionization response. 
In addition, the uncertainty on the $^{85}$Kr activity is combined with the spectral uncertainties from the $\beta$-decay Q-value and atomic exchange and screening effects~\cite{PhysRevA.90.012501,Haselschwardt:2020iey}, as discussed in~\refcite{DarkSide-50:2022qzh}. 

\begin{figure}[t]
    \centering
    \includegraphics[width=\linewidth]{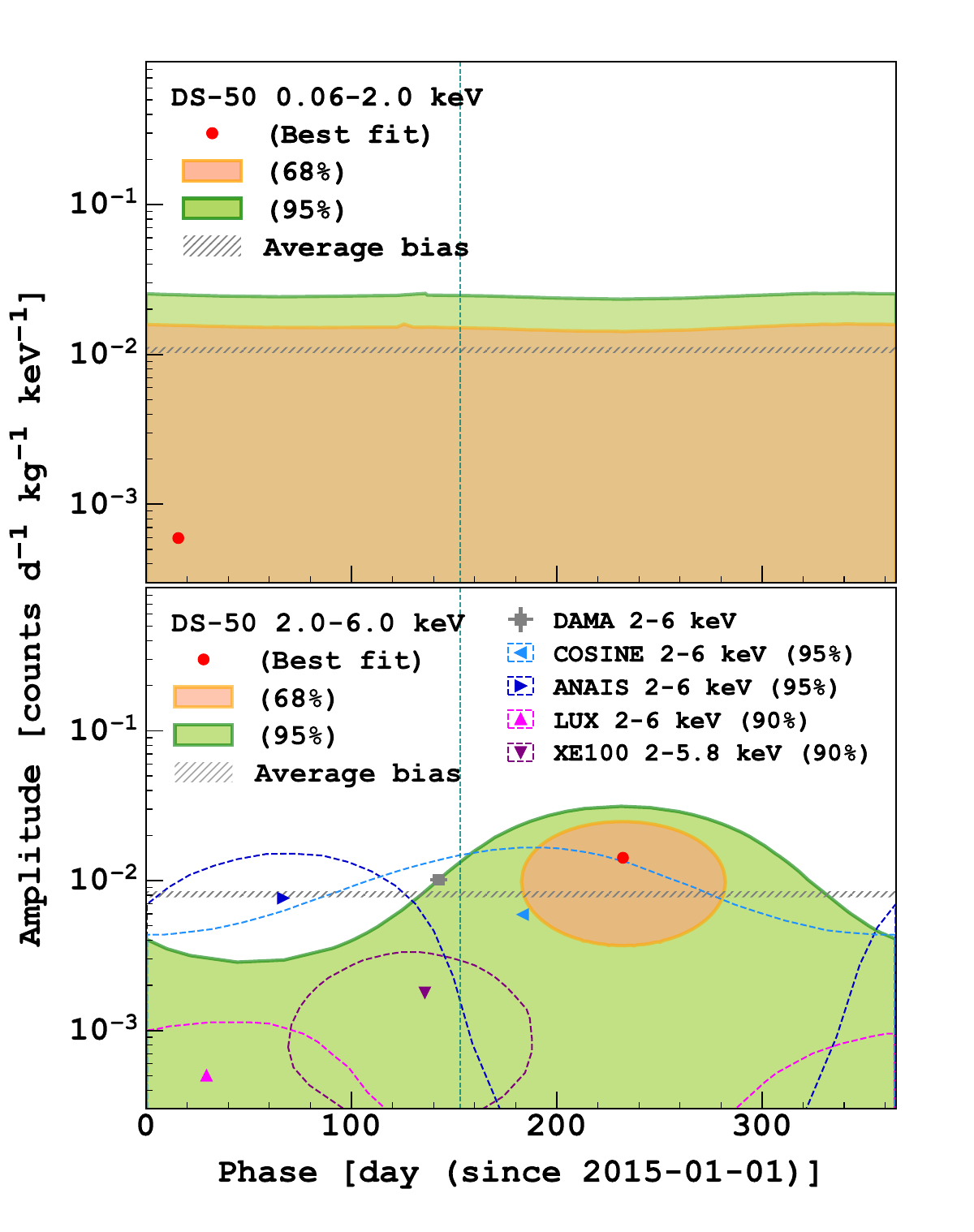}
    \caption{Best fit parameters in the phase versus amplitude space from the likelihood analysis with the fixed period of \SI{1}{\year}.
    The vertical dotted line represents the phase of the dark matter signal expected from the standard halo model.
    The horizontal dash-hatched line corresponds to the estimated biases in the fit, extracted from pseudo experiments.
    Also shown are the results from other experiments using NaI(Tl) crystal scintillators (DAMA/LIBRA~\cite{Bernabei:2021kdo}, COSINE-100~\cite{COSINE-100:2021zqh}, and ANAIS-112~\cite{Amare:2021yyu}) and liquid xenon TPC (XENON100~\cite{XENON:2017nik} and LUX~\cite{LUX:2018xvj}).}
    \label{fig:bestfit_amp_phase}
\end{figure}

\reffiginit{fig:bestfit_amp_phase} shows the best fit values of ($A_\chi$, $\phi$) when fitting the data with \refeqn{eq:likelihood}, together with the associated 68\% and 95\% confidence level~(CL) contours, for the two analyzed ranges. 
The $\chi^2/{\rm NDF}$~(number degrees of freedom) for the best-fit in [4, 41]~\si{\el} ([41, 68]~\si{\el}) is $132.6/124$ ($154.1/124$).
The same analysis has been repeated by varying the bin width from \SI{1}{\day} to \SI{10}{\day}, and no significant variations have been found.

\begin{figure}[ht]
    \centering
    \includegraphics[width=\linewidth]{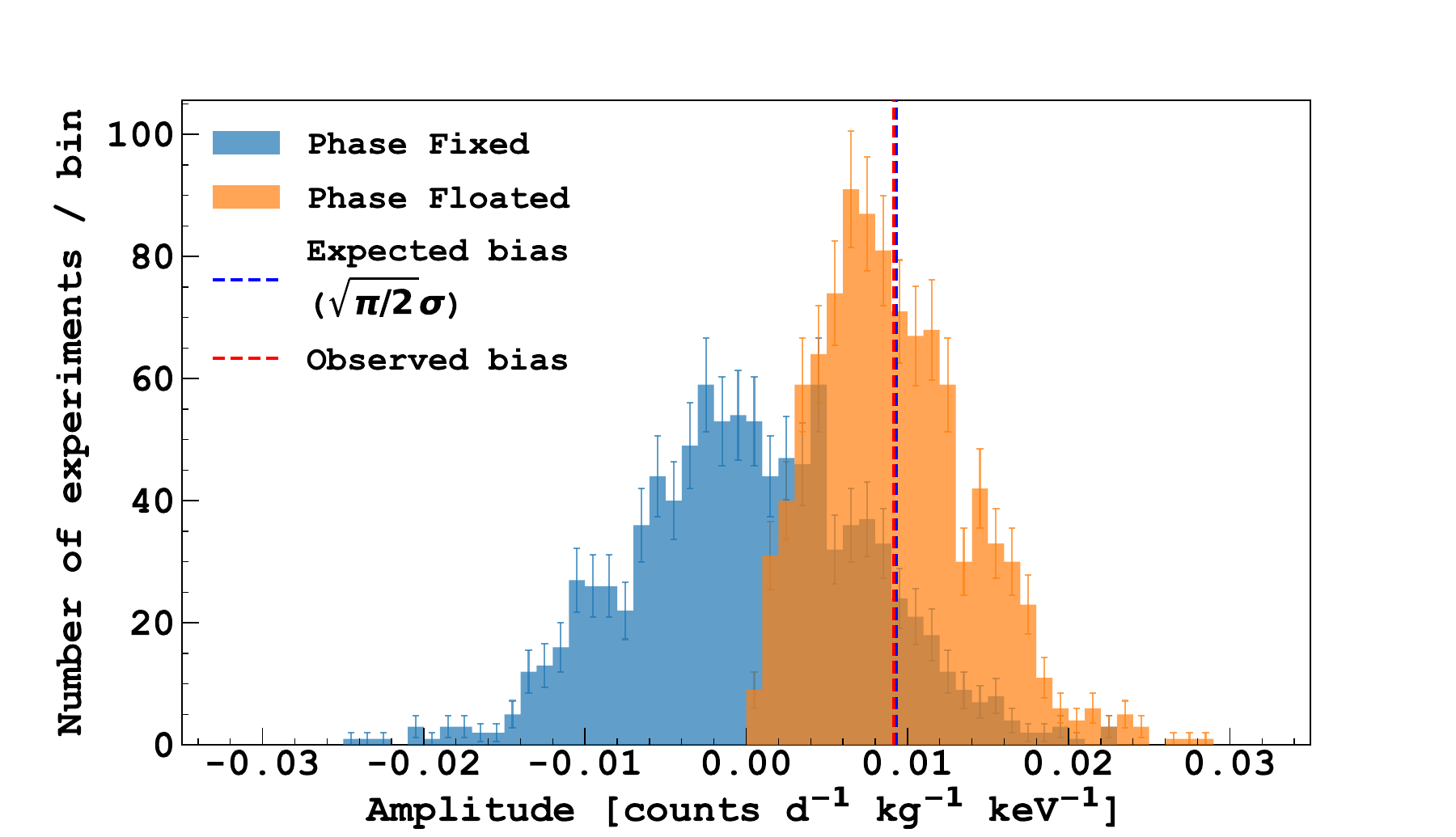}
    \caption{Distributions of the best-fit amplitude for background-only pseudo datasets.
    The vertical red line is the mean of the amplitude obtained by the fit, while the blue vertical line corresponds to $\sqrt{\pi/2}\sigma$ where $\sigma^2$ is the variance of the amplitude obtained by the fit fixing the phase.}
    \label{fig:bias}
\end{figure}

The fit does not show any evidence of modulation in either of the two energy ranges, however, it has to be noted here that the fit is expected to be biased due to the nonlinearity of the pair of the parameters of interest ($A_\chi$, $\phi$) to the signal rate in \refeqn{eq:pdf}\footnote{Although an unbiased fit could be performed by choosing the parameters as ($S_1$$\equiv$$A_\chi\cos\phi'$ , $S_2$$\equiv$$A_\chi\sin\phi'$), where $\phi'$$=$$\frac{\phi}{T/2\pi}$, we opt to use the nonlinear pair for the comparison to the other experiments.}~\cite{ALEGRIA2009748,Amare:2021yyu}.
We estimate the bias using 1000 pseudo experiments based on the background model without a modulation signal.
\reffiginit{fig:bias} shows an example of the result of such pseudo experiments in the [41,68]~\SI{}{\el} range.
The bias is extracted to be \SI{0.011}{counts\per\day\per\kilo\gram\per\keVee} (\SI{0.008}{counts\per\day\per\kilo\gram\per\keVee}) in the [4, 41]~\si{\el} ([41, 68]~\si{\el}) range as the mean of the amplitudes of the pseudo datasets.
It is confirmed that the result is consistent with the theoretically predicted bias, $\sqrt{\pi/2}\sigma(A_\chi)$, where $\sigma(A_\chi)$ is the variance of the modulation amplitude for a fixed phase.
The estimated bias is overlaid with a dash-hatched line in \reffig{fig:bestfit_amp_phase}.
The best fit results are consistent with the mean bias of the background-only pseudo samples within $1\sigma$.

The result in the [2.0, 6.0]~\si{\keVee} range is used to test the modulation observed by DAMA/LIBRA in the same interval, compatible with a dark matter signal over 14 cycles with a significance of \(>\)13\(\sigma\)~\cite{Bernabei:2021kdo}.
An assumption behind this analysis is that dark matter produces electron recoils in both NaI and LAr detectors with the same probability per unit detector mass
\footnote{For completeness, the [2.0,6.0]~\SI{}{\keVee} in this analysis corresponds to [8.6,21.6]~\SI{}{\keV} for nuclear recoils taking into account the quenching effect~\cite{DarkSide:2021bnz}.
Such energy range for nuclear recoils in turn corresponds to about [1,3]~\SI{}{\keVee} in NaI~\cite{Xu:2015wha,Lee:2024unz}.}.
The significance of the analysis is such that we can neither confirm nor reject the DAMA/LIBRA observation over the null hypothesis. 
For completeness, the same conclusion is drawn for the [1.0,3.0]~\SI{}{\keVee} range, also analyzed by DAMA/LIBRA.

\begin{table*}[t]
{\renewcommand{\arraystretch}{1.05}
{\setlength{\tabcolsep}{10pt}
\begin{threeparttable}
    \caption{List of the nuisance parameters, together with their central values (\(\theta^0_k\)) and uncertainties (\(\Delta \theta_k\)).
    The uncertainties are given as percentages of the corresponding central values.
    The uncertainties arising from the \(\beta\)-decay spectrum and the ionization response are reported in terms of the event rate.}
    \centering
    \begin{tabular}{c|c|c|c|c}
        \hline
        Parameter & Energy range & \(\theta^0_k\) & \(\Delta \theta_k\) & Refs.\\ \hline \hline
        T & \SI{1}{\year} & 0 \\ 
        Fiducial volume & all & \SI{19.4}{\kilo\gram} & 1.5\%\tnote{*} & \cite{DarkSide-50:2022qzh} \\ 
        \(\tau_\mathrm{^{37}Ar}\) & all & \SI{35.0}{\day} & 0 & \cite{be:cea-02476417} \\
        \(\tau_\mathrm{^{85}Kr}\) & all & \KrEightFiveHalfLife & 0 & \cite{be:cea-02476107} \\
        \(\tau_\mathrm{^{54}Mn}\) & all & \MnFiveFourHalfLife & 0 & \cite{be:cea-02476107} \\
        \(\tau_\mathrm{^{60}Co}\) & all & \CoSixZeroHalfLife & 0 & \cite{be:cea-02476243} \\ 
        \multirow{2}{*}{\(A_\mathrm{^{37}Ar}\)}  & [0.06, 2.0]~\si{\keVee} & \SI{0.85}{\ct\per\day\per\kilo\gram} & \multirow{2}{*}{14\%} & \\
        & [2.0, 6.0]~\si{\keVee} & \SI{2.1}{\ct\per\day\per\kilo\gram} & & \\
        \multirow{2}{*}{\(A_\mathrm{^{85}Kr}\)} & [0.06, 2.0]~\si{\keVee} & \SI{1.0}{\ct\per\day\per\kilo\gram} & \multirow{2}{*}{4.7\%} & \multirow{2}{*}{\cite{DarkSide-50:2022qzh}} \\
         & [2.0, 6.0]~\si{\keVee} & \SI{1.7}{\ct\per\day\per\kilo\gram} & & \\
        \multirow{2}{*}{\(A_\mathrm{^{54}Mn}\)} & [0.06, 2.0]~\si{\keVee} & \SI{0.01} {\ct\per\day\per\kilo\gram} & \multirow{2}{*}{40\%} & \multirow{2}{*}{\cite{DarkSide-50:2022qzh}} \\
        & [2.0, 6.0]~\si{\keVee} & \SI{0.02}{\ct\per\day\per\kilo\gram} & & \\
        \multirow{2}{*}{\(A_\mathrm{^{60}Co}\)} & [0.06, 2.0]~\si{\keVee} & \SI{0.25} {\ct\per\day\per\kilo\gram} & \multirow{2}{*}{12\%} & \multirow{2}{*}{\cite{DarkSide-50:2022qzh}} \\ 
        & [2.0, 6.0]~\si{\keVee} & \SI{0.58}{\ct\per\day\per\kilo\gram} & & \\
        \multirow{2}{*}{\isotope{85}{Kr} \(\beta\)-decay spectrum} & [0.06, 2.0]~\si{\keVee} & \SI{1.0}{\ct\per\day\per\kilo\gram} & 0.4\% & \multirow{2}{*}{\cite{PhysRevA.90.012501,Haselschwardt:2020iey,DarkSide-50:2022qzh}} \\
        & [2.0, 6.0]~\si{\keVee} & \SI{1.7}{\ct\per\day\per\kilo\gram} & 0.4\% & \\
        \multirow{2}{*}{Ionization response} & [0.06, 2.0]~\si{\keVee} & \SI{2.1}{\ct\per\day\per\kilo\gram} & 2.2\% & \multirow{2}{*}{\cite{DarkSide-50:2022qzh,DarkSide:2021bnz}} \\ 
        & [2.0, 6.0]~\si{\keVee} & \SI{4.4}{\ct\per\day\per\kilo\gram} & 0.1\% & \\ \hline
    \end{tabular}
    \begin{tablenotes}
        \item[*] More details in the text.
    \end{tablenotes}
    \label{tab:nuis_par}
\end{threeparttable}
}
}
\end{table*}

\subsection{Modulation amplitude as a function of energy}
Additional constraints on the modulation amplitude are obtained by simultaneously fitting the event timestamps and energies after fixing the period (\SI{1}{\year}) and the phase (maximum at June 2nd) to those expected from the Standard Halo Model~\cite{McCabe:2013kea,OHare:2019qxc}. 
This approach does not require any assumption on the SE rate and thus allows the range to be extended down to \SI{3}{\electron} or \SI{0.04}{\keVee}, which corresponds to the primary electron induced by the interaction plus, on average, two subsequent ionization electron.
The  likelihood,
\begin{multline}
  \mathcal{L}  =   \prod_{i\,\in\,t_\textrm{bins}} \prod_{j\,\in\,E\,_\textrm{bins}} \mathcal{P} \left( n_i^j| m_i^j(A_\chi^j,C^j,\tilde{\Theta})\right) \\
  \times  \prod_{\tilde{\theta_k}\,\in\, \tilde{\Theta}}\mathcal{G}(\tilde{\theta}_k|\tilde{\theta}^0_k\textrm{,}\,\Delta \tilde{\theta}_k), 
\label{eq:likelihood2}
\end{multline}
is the product of the Poisson probabilities in each of the $ij$-bins defined by the event time ($i$) and energy expressed in terms of number of electrons ($j$) given the signal amplitude, $A^j_\chi$, and the constant background component, $C^j$.
The chosen bin width along the time axis corresponds to \SI{7}{\day} and the bin widths along the energy axis are \SI{0.02}{\keVee} in the range $[0.04, 0.06]$~\si{\keVee}, \SI{0.25}{\keVee} in the range $[0.06, 1.0]$~\si{\keVee}, \SI{1}{\keVee} in the range $[1.0, 6.0]$ \si{\keVee}, and \SI{2}{\keVee} in the range $[6.0, 20.0]$ \si{\keVee}
The sample of events with \SI{3}{\el} ($[0.04, 0.06]$ \si{\keVee}) is contaminated by SE's.
To account for this background, we anchored its time variation to that of events below \SI{3}{\el}, selected in coincidence with the previous event, largely dominated by SE.
This approach is justified by the observation that the spectrum of events occurring in a \SI{2}{\milli\second} window from the previous event, which consists of more than 90\% of SE's, is stable over time.
The amplitude of the signal in each energy interval, $A^j_{\chi}$, is assumed uncorrelated with the others. 
Nuisance parameters $\tilde{\Theta}$, in \refeqn{eq:likelihood2} are the same as in \refeqn{eq:likelihood}, but account for energy spectral distortions of the background components as done in \refcite{DarkSide-50:2022qzh}.  

\begin{figure}[t]
    \centering
    \includegraphics[width=\linewidth]{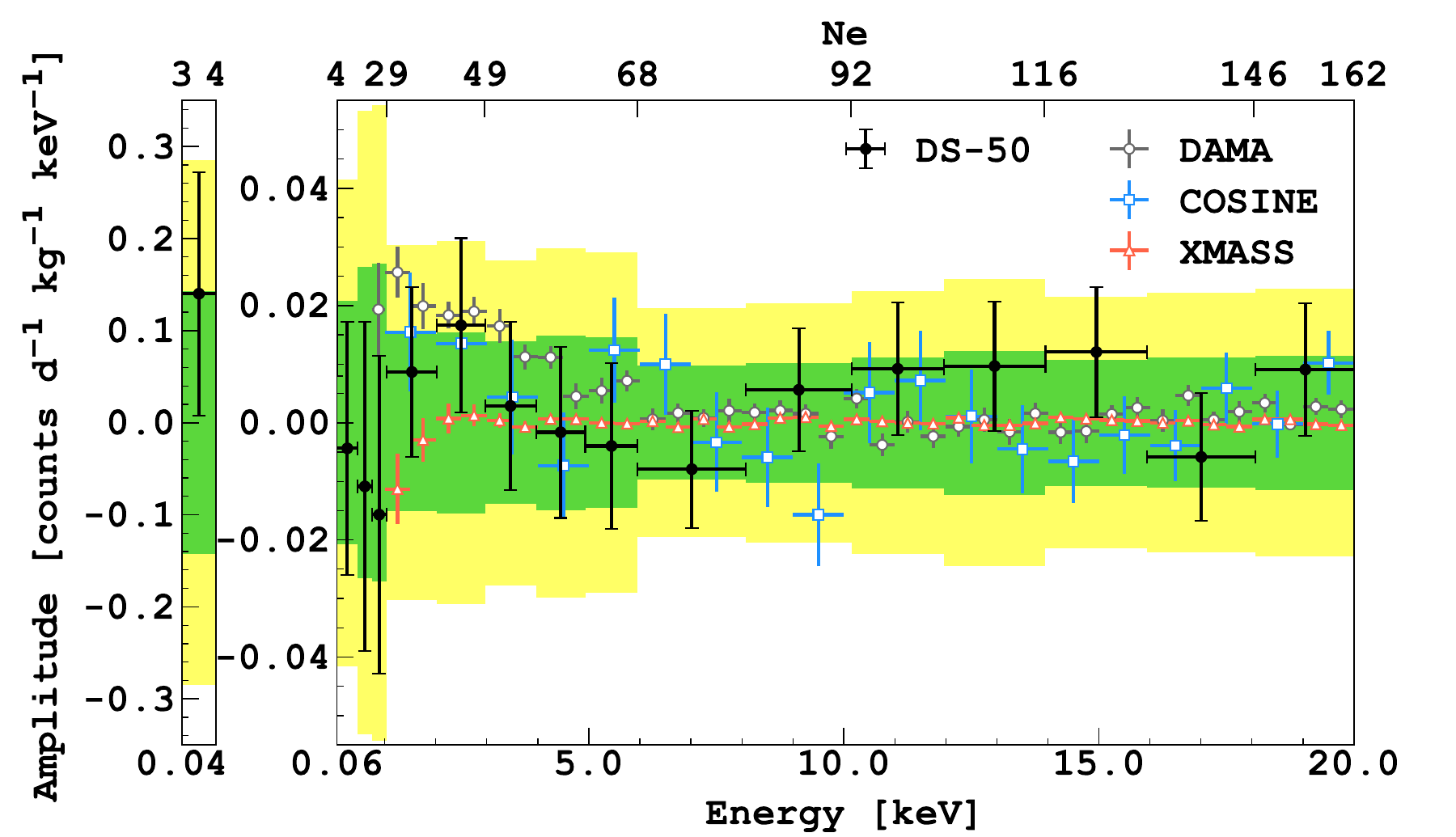}
    \caption{Best fit amplitude of the modulation signal as a function of \(N_e\).
    The green and yellow bands represent the expected \(1\sigma\) and \(2\sigma\) statistical fluctuations derived by background-only Monte Carlo samples.
    Also shown are the results from DAMA/LIBRA~\cite{Bernabei:2021kdo}, COSINE-100~\cite{COSINE-100:2021zqh}, and XMASS~\cite{XMASS:2022tkr}.}
    \label{fig:limit_amp}
\end{figure}

The measured event rate with \SI{3}{\el} is shown in \reffig{fig:rate_bestfit}, together with the fitted background model including the SE component.
\reffiginit{fig:limit_amp} shows the best-fitted amplitude as a function of the energy, together with the  1- and 2-$\sigma$ significance coverages, as derived with background-only Monte Carlo datasets.
The $\chi^2/{\rm NDF}$ for the best-fitted amplitude is $2275.9/2055$. 
The results from DAMA/LIBRA~\cite{Bernabei:2021kdo}, COSINE-100~\cite{COSINE-100:2021zqh}, and XMASS~\cite{XMASS:2022tkr} are also shown. 
In contrast to our approach, the DAMA/LIBRA looked at each energy bin independently and measured the amplitude by looking at the residuals of a yearly averaged event rate.

\subsection{Periodogram analysis}
\begin{figure}[t]
    \centering
    \includegraphics[width=\linewidth]{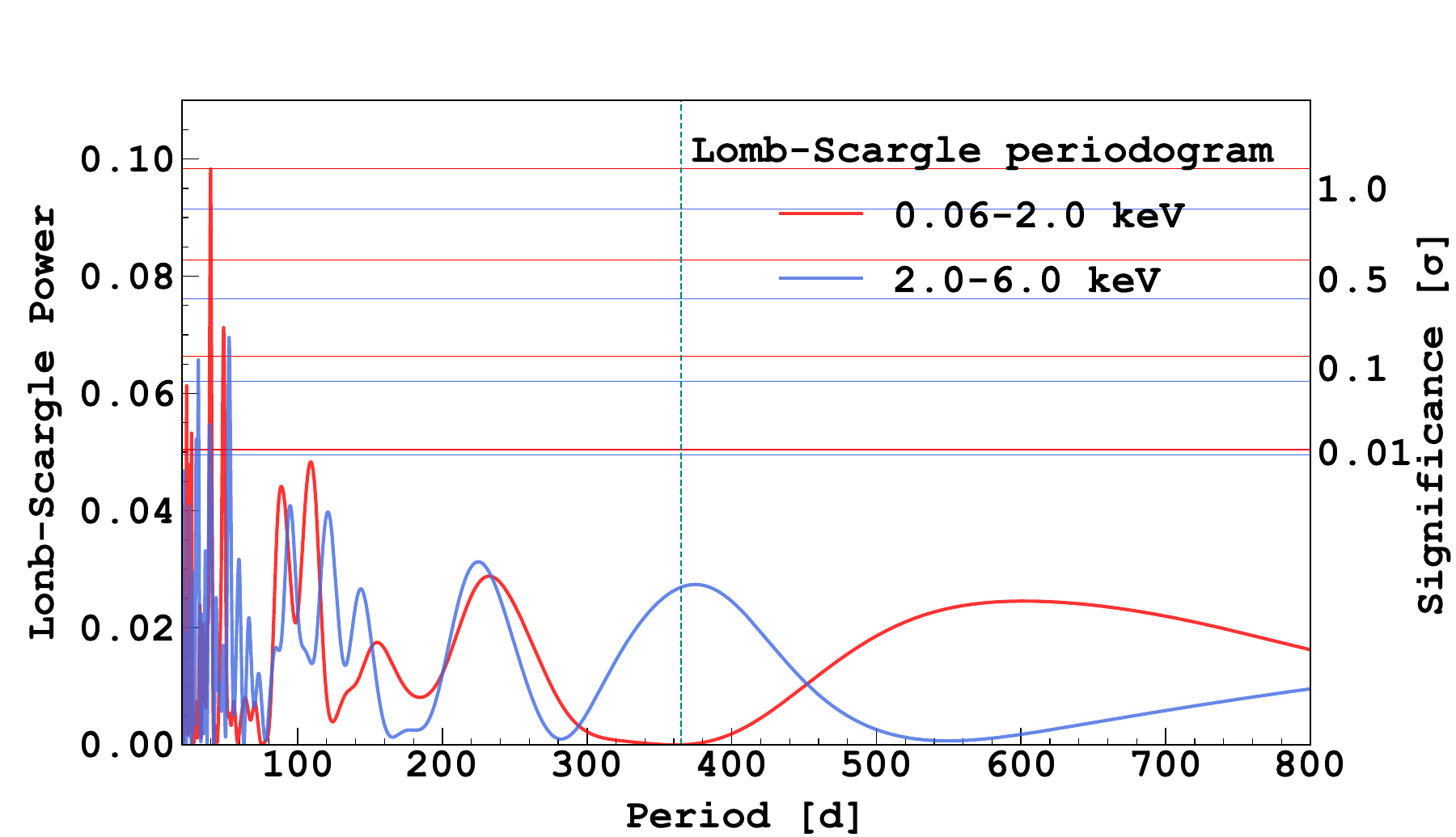}
    \caption{Observed sinusoidal signal strengths from the Lomb-Scargle periodogram as a function of its period. 
    The horizontal lines represent the $0.1\sigma$, $0.5\sigma$, and $1.0\sigma$ false alarm probability from the Bootstrap method for each range.
    The vertical dashed line corresponds to the period of \SI{1}{\year}.}
    \label{fig:ls}
\end{figure}

Finally, a Lomb-Scargle periodogram analysis is performed on the temporal evolution of the event rate to look for sinusoidal signals with any period, including the one expected from dark matter. 
The analysis is applied to the data residuals, after the subtraction of the best-fitted background model determined for each energy range independently from each other (i.e., \refeqn{eq:pdf} but $A_\chi$ is fixed to 0), as shown with the red and blue lines in \reffig{fig:rate_bestfit}. 
The uncertainty from the background fit is propagated to the data errors.
To assess the significance of the sinusoidal signals, we calculate the false alarm probability which is defined as the probability for a Gaussian noise background to produce a peak of the observed amplitude.
In this work, the bootstrap method~\cite{2018ApJS..236...16V} is adopted for the calculation.
The sensitivity of this analysis is evaluated by applying the Lomb-Scargle analysis over 1000 pseudo experiments where an annual modulation signal has been injected.
A median of 1\(\sigma\) significance for the false alarm probability is obtained with the addition of \SI{0.03}{\ct\per\day\per\kilo\gram\per\keV}.
The analysis of the data does not identify any significant modulation, scanning periods up to \SI{800}{\day}, as shown in~\reffig{fig:ls}.
For the period of \SI{1}{\year} for instance, the significance is lower than $0.01\sigma$ for both ranges.

%
\section{Conclusion}
We searched for an event rate modulation in the \DSf\ data between \num{0.06} and \SI{6.0}{\keVee} without assuming a specific dark matter signal model.
In none of the two analyzed ranges of $[0.06,2.0]$ \si{\keVee} and $[2.0, 6.0]$ \si{\keVee}, a modulation signal was observed within the sensitivity.
Also, a search is performed taking into account the background energy spectrum, which also failed to observe a significant modulation amplitude in the range [0.04, 20.0]~\si{\keVee}.
This is the first search for a dark matter-induced modulation signal in the sub-keV region.
Unfortunately, the significance of this result is not sufficient to confirm or reject the DAMA/LIBRA's positive observation in [0.75, 6.0]~\si{\keVee}.

The stability of the \DSf\ detector over nearly three years of operation, the accuracy of the background model, and the low-energy threshold achieved demonstrate the competitiveness of the dual-phase LAr-TPC technology in searching for modulation signals. 
This result is therefore promising in view of future massive dual-phase liquid argon experiments~\cite{GlobalArgonDarkMatter:2022ppc, DarkSide-20k:2017zyg, Franco:2015pha}, expected to reach much larger exposures and even lower background levels.

\begin{acknowledgements}
The DarkSide Collaboration offers its profound gratitude to the LNGS and its staff for their invaluable technical and logistical support. We also thank the Fermilab Particle Physics, Scientific, and Core Computing Divisions. Construction and operation of the DarkSide-50 detector was supported by the U.S. National Science Foundation (NSF) (Grants No. PHY-0919363, No. PHY-1004072, No. PHY-1004054, No. PHY-1242585, No. PHY-1314483, No. PHY-1314501, No. PHY-1314507, No. PHY-1352795, No. PHY-1622415, and associated collaborative grants No. PHY-1211308 and No. PHY-1455351), the Italian Istituto Nazionale di Fisica Nucleare, the U.S. Department of Energy (Contracts No. DE-FG02-91ER40671, No. DEAC02-07CH11359, and No. DE-AC05-76RL01830), the Polish NCN (Grant No. UMO-2019/33/B/ST2/02884) and the Polish Ministry for Education and Science (Grant No. 6811/IA/SP/2018). We also acknowledge financial support from the French Institut National de Physique Nucl\'eaire et de Physique des Particules (IN2P3),   the  IN2P3-COPIN consortium (Grant No. 20-152),  and the UnivEarthS LabEx program (Grants No. ANR-10-LABX-0023 and No. ANR-18-IDEX-0001),  from the São Paulo Research Foundation (FAPESP) (Grants No. 2016/09084-0 and No. 2021/11489-7),  from the Interdisciplinary Scientific and Educational School of Moscow University ``Fundamental and Applied Space Research'',  from the Program of the Ministry of Education and Science of the  Russian  Federation  for  higher  education  establishments,  project No. FZWG-2020-0032 (2019-1569), the International Research Agenda Programme AstroCeNT (MAB/2018/7) funded by the Foundation for Polish Science (FNP) from the European Regional Development Fund, and the European Union's Horizon 2020 research and innovation program under grant agreement No 952480 (DarkWave), the National Science Centre, Poland (2021/42/E/ST2/00331), and from the Science and Technology Facilities Council, United Kingdom.  I.~Albuquerque is partially supported by the Brazilian Research Council (CNPq). The theoretical calculation of beta decays was performed as part of the EMPIR Project 20FUN04 PrimA-LTD. This project has received funding from the EMPIR program co-financed by the Participating States and from the European Union’s Horizon 2020 research and innovation program. Isotopes used in this research were supplied by the United States Department of Energy Office of Science by the Isotope Program in the Office of Nuclear Physics.
 \end{acknowledgements}




\bibliography{ref}

\end{document}